\documentclass[12pt,preprint]{aastex}

\def\kms{\relax \ifmmode {\,\rm km\,s}^{-1}\else \,km\,s$^{-1}$\fi}

\def\Mso{{$M_{\odot}$}}

\def\cm3{${\rm cm}^{-3}~$}

\slugcomment{}

\shorttitle{}
\shortauthors{Villaver \& Livio}

\begin{document}

\title{The Orbital Evolution of Gas Giant Planets around Giant Stars}

\author{Eva Villaver}
\affil{Universidad Aut\'onoma de Madrid, Departamento de F\'{\i}sica
  Te\'orica C-XI, 28049 Madrid, Spain} 
\email{eva.villaver@uam.es}
\and
\author{Mario Livio}
\affil{Space Telescope Science Institute, 3700 San Martin Drive,
Baltimore, MD 21218, USA}
\email{mlivio@stsci.edu}

\begin{abstract}
Recent surveys have revealed a lack of close-in planets around evolved stars more massive than 1.2~$M_{\odot}$. Such planets are common around solar-mass stars. We have calculated the orbital evolution of planets around stars with a range of initial masses, and have shown how planetary orbits are affected by the evolution of the stars all the way to the tip of the Red Giant Branch (RGB). We find that tidal interaction can lead to the engulfment of close-in planets by evolved stars. The engulfment is more efficient for more-massive planets and less-massive stars. These results may explain the observed semi-major axis distribution of planets around evolved stars with masses larger than 1.5~$M_{\odot}$.

Our results also suggest that massive planets may form more efficiently around intermediate-mass stars.
\end{abstract}


\section{INTRODUCTION}

Observationally, due mostly to the inapplicability of high-precision Doppler 
techniques to stars with spectral types earlier than late-F (which have large rotational velocities and a small number of spectral lines), little was known until recently about the frequency of planets around the more-massive stars.  This situation has now changed, as surveys have been extended to  searches for planets around more-massive stars in an evolved stage \citep{Doli07,Doli09,Frin02,Hat03,Hat05,Hat06,Joh07a,Joh07b,Joh08,Liu07,Lm07,
Nied07,Nied09,Ref06,Sat07,Sat08,Set03,Set05}.  

One of the most important trends that these surveys have revealed is the lack
of close-in planets orbiting stars with masses $M > 1.5~M_{\sun}$ \citep{Joh07b,Sat08,Wri09} despite the fact that these planets are found around
$\approx$20 \% of the Main Sequence (MS) stars with $M < 1.2~M_{\sun}$. The
frequency of planets seems also to be higher around intermediate-mass stars
\citep{Lm07,Joh07b}. Furthermore, contrary to the correlation between metallicity and probability of planet-hosting found for solar-mass, main-sequence stars \citep[e.g.,][]{Fv05,San05}, the more-massive planet-hosting stars do not exhibit higher metallicities \citep[e.g.,][]{Pas07}.

On the theoretical side, it has been shown that the formation of Jupiter-mass planets around M~stars may be hindered \citep[e.g.,][]{L04,Il05}, while the probability that a given star has at least one gas giant increases linearly with the stellar mass up of 3~$M_{\odot}$ \citep[e.g.,][]{Kk08}.

One possibility is that the observed difference in the orbital distribution of planets found around intermediate and solar-mass primaries is due to the evolution of the star. Since all of the planets orbiting red giants or subgiants with $M > 1.5~M_{\sun}$ have semi-major axes $a > 0.5$~AU it has been suggested 
that the planets might be engulfed as the star evolves off the MS
\citep{Joh07b,Sat08}. This possibility is, however, often dismissed out in the literature with the argument that high-mass stars are physically too small to engulf hot Jupiters \citep{Joh07b,Joh08,Cur09}. Another possibility is that the observed differences in orbital distribution are primordial, and and they are a consequence of the planet formation mechanism around more massive stars.  Along these lines, it has been shown by \cite{Cur09} that the dependence of the lifetime of the gaseous disk on the stellar mass could result in halting the inward migration of planets around high-mass stars,  thus explaining the observed lack of short period planets around these stars. 

The point we are making in the present paper is that before stellar evolution can be ruled out as the mechanism behind the observed semi-major axis distribution of planets around evolved stars, detailed modeling of the orbital evolution needs to be performed. In other words, in order to determine the potential role of the stellar mass in the planet formation process, the effects of the evolution of the star on the observed orbit distribution around giants have to be correctly isolated. This is precisely the goal of this paper. 

\section{THE EQUATIONS}
There are several competing processes that affect the orbital distance
between the star and the planet as the star evolves off the MS: the changes in the mass of both the planet and the star, the gravitational and frictional drag, and the tidal force. 

To determine the rate of change in the planet's mass, we consider a planet of mass $M_p$ and radius $R_p$ moving with a velocity, $v$,  in a  circular orbit ($e = 0$) around a star of mass $M_*$. Since the planet is moving supersonically through the matter ejected by the giant star, it accretes mass. The accretion rate onto the planet, $\dot{M_p}\,_\mathrm{|acc}$, is given approximately by the Bondi-Hoyle expression \citep[e.g.,][]{Bh44, Rs71}, 
\begin{equation}
\dot{M_p}\,_\mathrm{|acc} = \pi R_A^2 \rho v~~,
\label{bondi}
\end{equation}
where $\rho$ is the density of the environment and $R_A$ is the accretion radius ($R_A= 2G M_p/v^2$ with $G$ the gravitational constant).  At very short distances (where $R_A\la R_p$) we have replaced $R^2_A$ by $R_AR_p$ to correct the geometrical radius by gravitational focusing effects.

At the same time, the planet's surface is being heated by radiation arising
from the shock front and from the stellar surface. This heating can lead to evaporation of surface  material. We estimate the evaporation rate, $\dot{M_p}\,_\mathrm{|ev}$ as in Villaver \& Livio (2007; Eq.~9). The temperature at the planet's surface has been taken to be the maximum between the radiative equilibrium temperature of the planet (see e.g., Eq.~5 in \citealt{Vl07}) and  the temperature of the shocked gas $T_{sh} = (3 m_H/16 k_b) v^2$  (estimated from the Rankine-Hugoniot conditions for an adiabatic shock where $k_b$ is Boltzman's constant and $m_h$ the mass of the hydrogen atom).  

The rate of change in the planet mass is thus given by
\begin{equation}
\dot{M_p}=(\dot{M_p}\,_\mathrm{|acc}-\dot{M_p}\,_\mathrm{|ev})~~.
\end{equation}

The rate of change of the stellar mass is simply 
$\dot{M_*}= - \dot{M}_\mathrm{mlr}$, where $\dot{M}_\mathrm{mlr}$ is the stellar mass-loss rate. Using Reimers' law for Red Giants \citep{Rei75},
\begin{equation}
\dot{M}_\mathrm{mlr} = 4 \times 10^{-13}\, \eta_R \frac{L_* R_*}{M_*} \quad [M_{\sun}~\mathrm{yr}^{-1}]~~,
\end{equation}
where $L_*$, $R_*$ and $M_*$ are the stellar luminosity, radius, and mass  respectively (in solar units) and $\eta_R$ is the Reimers parameter, which we take throughout this work to be $\eta_R= 0.6$
 
Conservation of angular momentum gives the equation for the rate of change in the orbital radius of the planet (see, e.g., \citealt{Ale76,Ls84}), 
\begin{equation} 
\left(\frac{\dot{a}}{a}\right) =-\frac{\dot{M_*}+\dot{M_p}}{M_*+M_p}-\frac{2}{M_p v}
\left[F_f+F_g\right]-\left(\frac{\dot{a}}{a}\right)_{t}~~,
\label{todo}
\end{equation}
where $F_f$ and $F_g$ are respectively the frictional and gravitational drag
forces and $({\dot{a}}/{a})_{t}$ is the rate of orbital decay due to the tidal interaction.

The gravitational drag force, $F_g$, arises from the eddying motions that are set up in the fluid by the passage of the planet. It is a consequence of the gravitational interaction of the planet with a gaseous medium. The drag force is given by \citep[e.g.,][and references therein]{Ost99}
\begin{equation} 
F_g= 4\pi \frac{(G M_p)^2 }{c_s^2}\rho I~~,
\label{fg}
\end{equation}
where $I$ is a time-dependent function of the Mach number. The numerical results of \citet{Ost99} show that for the Mach numbers encountered here, $I$ is approximately constant and has the value $I\simeq0.5$.  

The loss of angular momentum associated to the frictional force $F_f$ is proportional to the surface area of the planet exposed to the flow and it can be expressed in the form \citep[e.g.,][]{Rosen63} 
\begin{equation}
F_f = \frac{1}{2} C_d \rho v^2 \pi R_p^2~~,
\label{ff}
\end{equation}
where $C_d\simeq0.9$ is the dimensionless drag coefficient for a sphere.

Finally, the angular momentum loss associated to the tidal term
$({\dot{a}}/{a})_{t}$ arises from the additional force (besides the gravitational pull between the two centers of mass) resulting from the non-spherical part of the mass distribution from the tidally distorted
companion. In giant stars, which have massive convective envelopes, the most efficient mechanism to produce tidal friction is turbulent viscosity
\citep[e.g.,][]{Zah66,Zah77,Zah89}. The dissipation timescale is determined by the effective eddy viscosity, with eddy velocities and length scales given approximately by standard mixing length theory if convection transports most of the energy flux \citep{Zah89,Vp95,Retal96}. The tidal term is given by 
\begin{equation}
\left(\frac{\dot{a}}{a}\right)_{t} = \frac{f}{\tau_d} \frac{M_\mathrm{env}}{M_*} q (1+q)
\left(\frac{R_*}{a}\right)^8~~,
\label{tidal}
\end{equation}
with $M_\mathrm{env}$ being the mass in the convective envelope, $q = M_p/M_*$, and $\tau_d$ the eddy turnover timescale, given in the case of a convective envelope \citep{Retal96},  
\begin{equation}
 \tau_d = \left[\frac{M_\mathrm{env} (R_*-R_\mathrm{env})^2 }{3L_*}\right]^{1/3}~~,
\label{eddy}
\end{equation}
where $R_\mathrm{env}$ is the radius at the base of the convective envelope. The term $f$ in Eq.~(\ref{tidal}) is a numerical factor obtained from integrating the viscous dissipation of the tidal energy across the convective zone. \cite{Zah89} used $f=1.01(\alpha/2)$ where $\alpha$ is the mixing length parameter. \cite{Vp95} confirmed that observations are consistent with $f\approx$1 (specifically, they obtained values in the range $0.5 \lesssim f \lesssim 2$ with a preference to values in the upper half of the range) as long as $\tau_d \ll P$ with $P$ being the orbital period. We therefore used $f = (P / 2\tau_d)^2$ to account only for the convective cells that can contribute to viscosity when  $\tau_d > P/2$, otherwise we take $f=1$.

The evolution of orbital eccentricities is beyond the scope of the present paper. The damping of eccentricities due to tidal forces may eventually lead to a  narrow distribution of e-values for close-in planets \citep[e.g.,][ ]{Retal96,Jac08a}. Note that the initial value of the eccentricity has little effect on the orbital decay rate \citep[e.g.,][]{Jac08b}.

Calculation of the tidal term requires a knowledge of the structure of the star. We have used detailed stellar models provided to us by Lionel Siess. These were calculated  based on the stellar evolution code STAREVOL described in \cite{Siess06}.  We have used stars with MS masses of 1, 2, 3, and 5 \Mso, solar metallicity, and a mass-loss prescription with a Reimers parameter of $\eta_R= 0.6$.

\section{RESULTS}
To determine the evolution of the planet's orbit we integrate Eq.~\ref{todo}
along the RGB. The stellar evolution timescales (in the absence of significant mass loss) are set by the rate of consumption of the nuclear fuel. Since the nuclear burning has been exhausted in the core, during the RGB hydrogen burning continues in a shell outside the helium core, which now, devoid of energy sources, is contracting and heating up. As the core contracts the envelope expands and cools. The stellar effective temperature decreases while the star's radius and luminosity increase.  We have updated the stellar parameters at each time step of the orbital evolution.

As the star evolves off the MS the wind velocity decreases ($V_\mathrm{wind}\approx 5$--10~\kms). The termination shock which marks the position where the stellar wind interacts with the ISM cools down and the temperature of the medium at the planet's location has an upper limit provided by the stellar effective temperature \citep[e.g.,][]{Vgm02}. We used $T = 2500$~K as the temperature of the ambient medium. The Mach number $M = v/c_s$ is then always $M>3$ for the orbital distances relevant to this work. 

The density of the environment has been computed as
\begin{equation}
\rho = \frac{\dot{M}_\mathrm{mlr}}{4 \pi a^2 V_\mathrm{wind}}~~,
\end{equation}
unless $\dot{M}_\mathrm{mlr} = 0$, in which case we took a number density of $\rho=0.001$~cm$^{-3}$. We use a wind velocity  of 5~\kms along the RGB.

\begin{figure}
\plotone{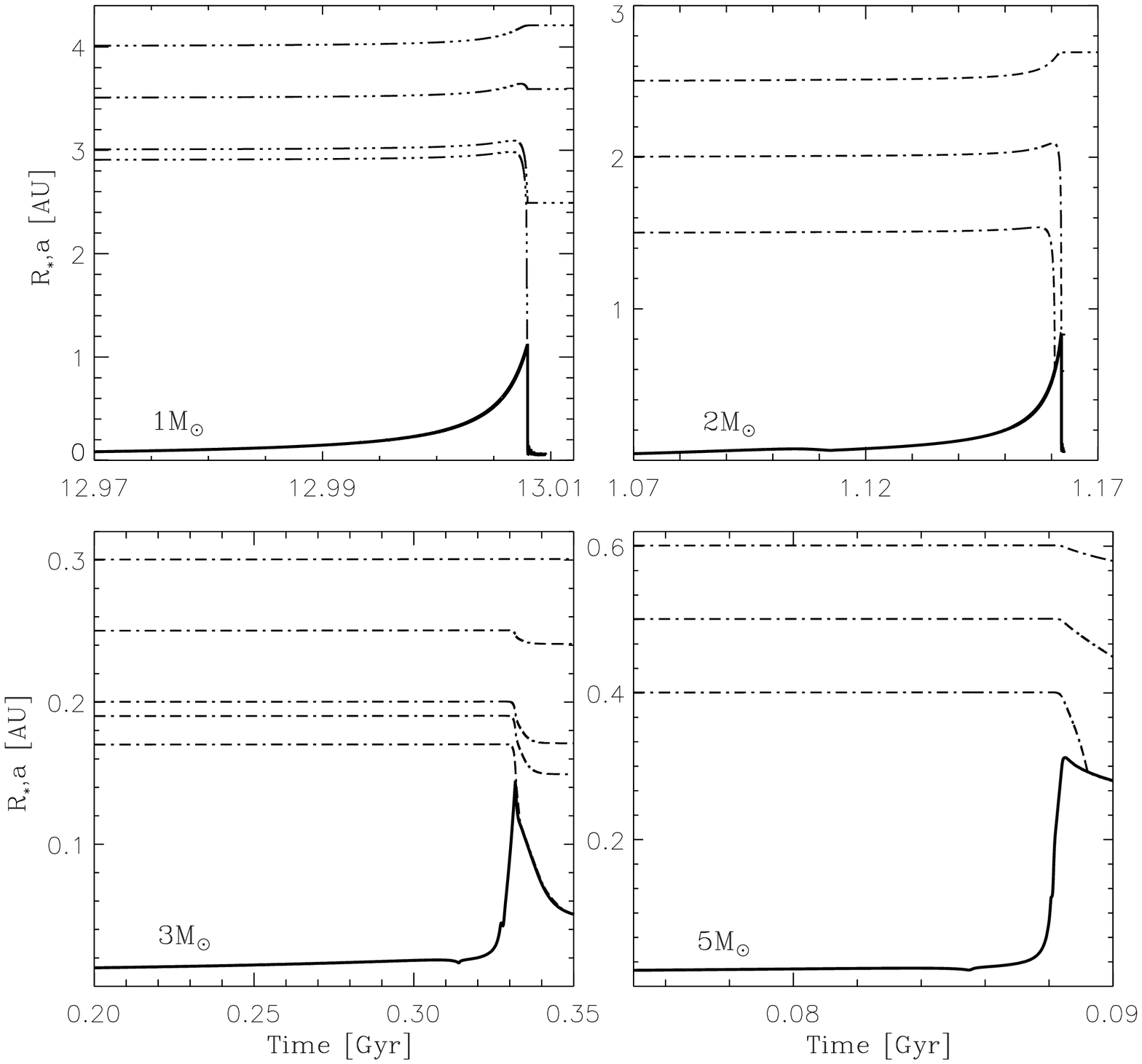}
\caption[ ]{Evolution of the orbital separation of a planet with $M_p = M_{J}$ (dash-dotted line) and the radius of the star (solid line) along the RGB. The different panels represent the stellar MS masses considered (1, 2, 3 and 5~$M_{\sun}$ from left to right and top to bottom respectively) and are marked at the bottom left corner of each plot. The different dashed lines are the orbits for different initial orbital distances.
\label{fig_jupiters}}
\end{figure}

We have computed the orbital evolution for a range of initial orbital
distances and planet and stellar masses in order to determine the minimum
initial orbital distance for which a planet will avoid being tidally captured by the expanding star. For illustration, some of the computed orbits (dash-dotted lines) are plotted in Fig.~\ref{fig_jupiters}, where we have used a planet with a mass of $M_p = 1$ $M_J$ orbiting stars with different MS masses to the end of the RGB phase. The evolution of the stellar radius is shown as a solid line. The panels, from top to bottom and left to right, are for stars with MS masses of 1, 2, 3, and 5~$M_{\sun}$ respectively. Note that the scale on the axes is different for each panel. 

At large distances from the star, the densities involved are low, and the drag terms associated with the forces F$_f$ and F$_g$ in Eq.~(\ref{todo}) play a negligible role in the evolution of the orbit. Moreover, since the accretion rate onto the planet is always small compared to the stellar mass-loss rate, the first term in Eq.~(\ref{todo}) is dominated by $\dot{M_*}$. The temporal behavior of the orbit is then mostly governed by the relative
importance of the terms associated with the stellar mass-loss $\dot{M_*}$ and the tidal interaction $(\dot{a}/a)_t$.  
 
Red giant mass-loss rates are somewhat uncertain. As noted in \S2 we have in the present work estimated the mass-loss rate by the Reimers prescription with $\eta_R= 0.6$. This seems to reproduce fairly well the observations of individual RGB stars. The peak RGB mass-loss rates are higher for lower-mass stars (10$^{-8}$ for 1~$M_\sun$ versus 10$^{-10}$ for the 3~$M_\sun$), and the lowest mass stars also reach the largest radius at the tip of the RGB. 

Figure~\ref{fig_jupiters} demonstrates the three possible outcomes of orbital evolution:  (i)~Beyond a certain initial orbital separation, the orbital separation simply increases, due to systemic mass loss. (ii)~There is a range of initial orbital separations for which the orbit decays, but the planet avoids being engulfed. (iii)~Inward from some critical, initial orbital separation, the planet is engulfed mostly due to tidal interaction.   

With the purpose of quantifying the influence of the planet's mass, we have also integrated the orbit of planets with masses of 3 and 5~$M_{J}$ (see Figs.~\ref{fig_2m} and \ref{fig_3m}). The top panel of Fig.~\ref{fig_2m} shows the RGB evolution of a 2~$M_\sun$ star along the HR diagram (Fig.~\ref{fig_3m} is the same, but for a 3~$M_\sun$ star). The bottom panel shows the orbital evolution for planets with masses 3~$M_J$ (dotted line)
and 5~$M_J$ (dash-dotted line) together with the evolution of the stellar
radius (solid line). Note that the bottom panel does not show the maximum 
extent of the stellar radius, but a zoom of a section at the base of RGB, to better appreciate the details. We selected small initial orbits to identify at which points during the RGB these planets are swallowed by their stars (marked by the location of the arrows). 

For all the initial orbits that satisfy the condition $a_o \leq R_*^\mathrm{max}$ the planet gets engulfed by the star at same point before the end of the RGB phase. The more massive the planet the stronger is the tidal interaction with the star, and therefore the sooner the orbit decays to meet the stellar radius (see Figs.~\ref{fig_2m} and \ref{fig_3m}). 

Some of our findings are summarized in Table~\ref{tableorbit}, where we
list the minimum initial orbital distance for which a planet with a given
mass avoids being engulfed by the star.  The second column gives the maximum radius reached by the star on the RGB, $R_*^\mathrm{max}$ (in AU), and the following columns give the minimum orbital distances (in AU) at which planets with masses of 1, 3, and 5~M$_{J}$ (respectively)  avoid being engulfed. For the stellar masses of 1 and 2~$M_{\sun}$, we calculated a grid using initial orbits at steps of 0.1 AU, for the  3 and 5~$M_{\sun}$ stars we used initial orbits at steps of 0.01 and 0.05 AU,  respectively. 
 
Two important conclusions can be extracted from Table~\ref{tableorbit}. First, the tidal ``capture'' radius increases with the planet's mass; a Jupiter-mass planet is captured by a 2~$M_{\sun}$ star if it starts at an initial orbit of 2.1~AU while a 5~$M_J$ planet will be engulfed by the star if it has an initial orbit $a_o < 2.5$~AU. Second, the tidal capture radius decreases with increasing stellar mass; for a 5~$M_J$ planet the initial orbit has to be larger than $a_o \approx 3 \times R_*^\mathrm{max}$ to avoid tidal capture around a 2~$M_\sun$ star, while it has to be larger than $a_o \approx 2 \times R_*^\mathrm{max}$ to avoid tidal capture around a 5~$M_\sun$ star.

\section{DISCUSSION AND CONCLUSIONS}

Our main goal has been to determine whether stellar evolution could explain the observed distribution of the semi-major axes of planetary orbits  around evolved stars (i.e., semi-major axis $>$\,0.5~AU). We found that when the details of the orbital evolution are accurately calculated, tidal interactions constitute a quite powerful mechanism, capable of capturing 
close-in planets into the envelope of evolved stars.

\begin{figure}
\plotone{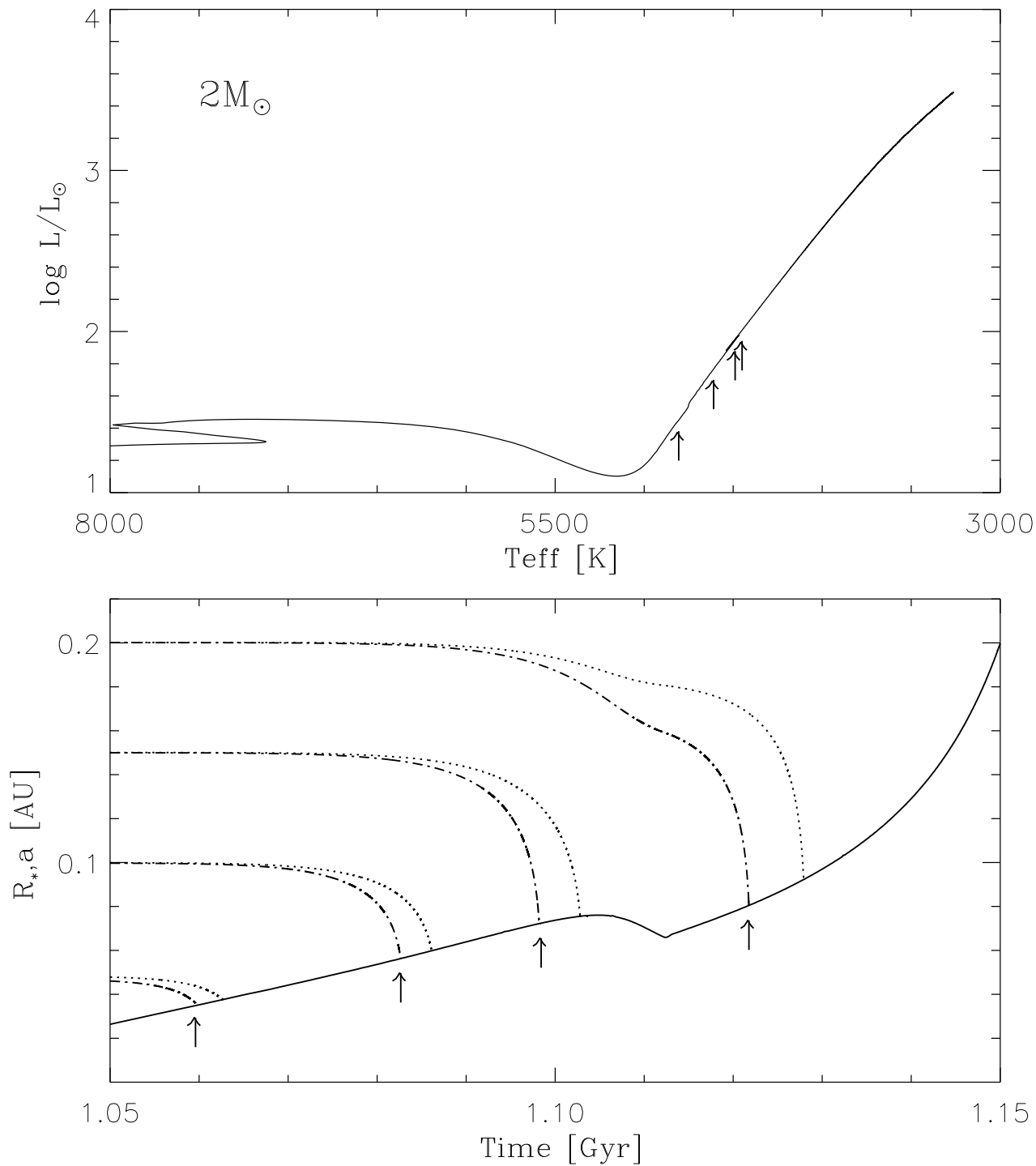}
\caption[ ]{Top: The evolution of a 2~$M_{\sun}$ star during the RGB along the HR diagram.  Bottom: The evolution of the orbital separation of a planet with $M_p = 3~M_{J}$ (dotted line) and $M_p = 5~M_{J}$ (dash-dotted line) for four different initial orbital separations. The evolution of the stellar radius along the RGB is also shown as a solid line. The arrows show the location at which the $M_p = 5~M_{J}$ planet enters the stellar envelope for the different initial orbits. Note that the maximum radius reached by the star during the RGB has not been plotted in the bottom panel.
\label{fig_2m}}
\end{figure}

\begin{figure}
\plotone{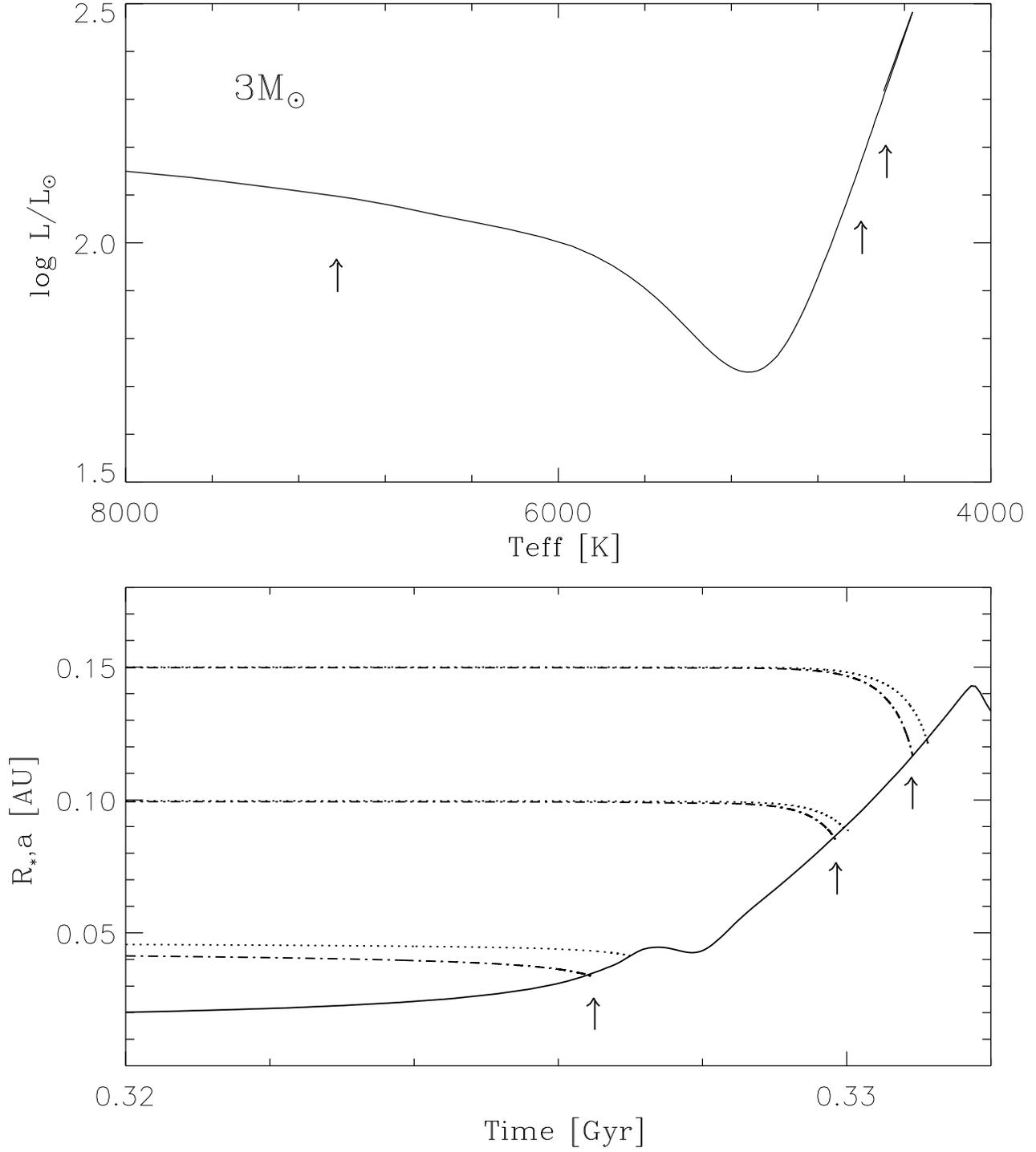}
\caption[ ]{Same as Fig.~\ref{fig_2m} but for a star with a MS mass of 3~$M_\sun$. In the bottom panel the star reaches the RGB maximum radius. 
\label{fig_3m}}
\end{figure}

To date, there are $\sim$20 exoplanets discovered around giant stars with $M
> 1.5~M_{\sun}$. The host stars have radii in the range $0.02 < R_* < 0.1$~AU. Our models are consistent with the existence of these planets at the point in the RGB evolution at which they are observed (see e.g., Table~8 in \citealt{Sat08} and our Figs.~\ref{fig_2m} and~\ref{fig_3m}).  However, we do not expect to find massive planets with a $<$\,0.4~AU around a 2~$M_\sun$ star with a R$_*\geq 0.1$~AU (or 24~$R_\sun$). Our calculations  provide the minimum orbital radius inside of which planets will be engulfed by the star at the end of the RGB evolution.

We find that the evolution of the star alone can quantitatively explain the observed lack of close-in planets around evolved stars even allowing for the uncertainties associated with mechanisms such as mass loss along the RGB or tidal-interaction theory. A mechanism such as the one invoked by Currie (2009; i.e., a lifetime stellar-mass dependency of the gas in the planet-forming disk that can halt migration) is not needed, although it might still be present.

We find that given an initial distance at which tidal capture is possible, the
more massive the planet, the earlier it will be captured by the RGB envelope. Observationally, it appears that giant stars host more massive planets than MS stars \citep[e.g.,][]{Joh07b,Lm07}. Since we find that more massive planets are expected to be engulfed earlier in the RGB evolution, the fact that they are more frequently observed may point towards a planet-formation mechanism that favors the formation of more-massive planets around intermediate-mass stars. This would be consistent with a scenario in which the disk mass scales with the stellar mass, and more massive disks produce more massive planets (see also \citealt{Kk08}).

Along similar lines, since we find a high probability of tidal capture of the planet by evolved stars, the higher frequency of planets observed around intermediate-mass stars \citep{Lm07,Joh07b} seems to imply that the
efficiency of planet formation must be considerably higher for more massive stars, compared to their solar analogous.  

\cite{Ass09} estimated the probability of detecting transits of planetary
companions to giant stars to be $\geq$ 10 \% for several of the known
systems. Since tidal oribital decay decreases the initial orbital distance, it may increase the probability for the planet to be observed in transit. 

Our results suggest that many planets may be accreted by their host star \citep[see also][]{SL99a,SL99b}. Although so far the results are based on a fairly limited sample, it appears that giant stars hosting planets have the same metallicity distribution as giant stars without planets \citep{Pas07}. On the other hand, we should note that if a giant star engulfs a close-in planet early in the RGB, it will appear to be a giant star without a planet. This mechanism might perhaps provide for a partial explanation for the lack of correlation with metallicity.

Finally, stars that have accreted planets may show a higher spin rate, as the planet's orbital angular momentum is transferred to the star \citep[e.g.][]{LS02}. This phenomenon has been investigated by \citet{Mass08} and \citet{Carl09}, where they estimate the probability of finding rapid rotators among evolved stars. Our quantification of the tidal capture radius may help refine these calculations.

\acknowledgments 
We are very grateful to Lionel Siess, who used his stellar evolution code to calculate the stellar models used in the calculations of the orbital evolution.

\clearpage

\begin{deluxetable}{ccccc}
\tabletypesize{}
\tablecaption{Minimum Orbital Radius to Avoid Tidal Capture}
\tablewidth{0pt}
\tablehead{
\colhead{$M_*$} &\colhead{$R_*^\mathrm{max}$ [AU]} 
&\multicolumn{3 } {c}{ a$_\mathrm{min}$ [AU]}\\
&&
\colhead{$M_p = M_{J}$} &
\colhead{$M_p = 3~M_{J}$}&
\colhead{$M_p = 5~M_{J}$}
}
\startdata
1 $M_{\sun}$ &1.10 &3.00 &3.40 &3.70\\
2 $M_{\sun}$ &0.84 &2.10 &2.40 &2.50\\
3 $M_{\sun}$ &0.14 &0.18 &0.23 &0.25\\
5 $M_{\sun}$ &0.31 &0.45 &0.55 &0.60\\
\enddata
\label{tableorbit}
\end{deluxetable}

\end{document}